\documentclass[aps,12pt,nofootinbib,showpacs,showkeys,preprintnumbers,amsmath,amssymb]{revtex4-1}		
\usepackage{amsmath,amsfonts,amssymb,amsthm,bm,bbm,bbold,braket,color,dsfont,fancybox,hyperref,
srcltx,subcaption,graphicx,ucs,yfonts,cancel,xfrac,verbatim,xcolor,slashed}

\usepackage{float}
\usepackage{tikz-feynman}

\usepackage[section]{placeins}

\usepackage{lineno}

\begin{document}

\title{Introducing Model Benchmark Analysis for dark matter models with isomorphic lagrangians}

\author{Paulo Areyuna C.$^{1,2}$}
\email{paulo.areyuna@sansano.usm.cl}

\author{Jilberto Zamora-Saa$^{1,3}$}
\email{jilberto.zamora@unab.cl} 

\author{Alfonso R. Zerwekh$^{1,4,5}$}
\email{alfonso.zerwekh@usm.cl}

\affiliation{$^1$Millennium Institute for Subatomic physics at high
energy frontier - SAPHIR, Fernandez Concha 700, Santiago, Chile.}
\affiliation{$^2$Departamento de Física, Facultad de Ciencias, Universidad de La Serena, Avenida Cisternas
1200, La Serena, Chile}
\affiliation{$^3$Center for Theoretical and Experimental Particle Physics - CTEPP, Facultad de Ciencias Exactas, Universidad Andres Bello, Fernandez Concha 700, Santiago, Chile.}
\affiliation{$^4$Departamento de Física, Universidad Técnica Federico Santa María Casilla 110-V, Valparaíso, Chile.}
\affiliation{$^5$Centro Científico - Tecnológico de Valparaíso, Casilla 110-V, Valparaíso, Chile}

\begin{abstract}

In this work, we propose a novel approach to particle physics phenomenology, which we dubbed as model benchmark analysis. We consider models that provide vertices with similar topology but different Lorentz and gauge structure, which we call isomorphic models. Under this setup, the coupling constants for each theory have the same meaning, in the sense that describe the topologically equivalent diagrams. Consequently,
each model acts like a map of parameter space into the observable space. This definition allows to group isomorphic models into families with a common parameter space and perform calculations for each model. As a proof of concept, we obtained some guidelines for the search of leptophilic dark fermions at CLIC, making predictions oriented to discriminate between beyond the standard model scenarios in light of possible signals of new physics that could appear at lepton colliders. We present different observables that can be useful to help discriminating between models.
\end{abstract}
\keywords{Heavy Neutral Leptons, Scotogenic Model, Dark Matter, Lepton Collider, CLIC}
\maketitle

\section{Introduction}
While the standard model of particle physics (SM) has been highly successful in explaining fundamental interactions, this theory presents limitations for explaining certain phenomena, such as the Dark Matter problem (DM) and neutrino mass generation. In order to solve these problems, a significant part of work performed in theoretical particle physics has been devoted to creating and studying theories beyond the standard model (BSM). Despite the major efforts performed by the experimental particle physics community, there are no significant hints of physics beyond the standard model, making an imbalance between the number of models and the evidence to support them. Even under the assumption that a BSM signal can be measured in the near future, the vast number and everyday increasing number of new models that are proposed every day makes it difficult to make a connection between an excess of events and a specific particle physics model, specially for the dark matter problem, where the usual signature at colliders is missing energy (we can say nothing about something invisible). In light of this problem, different approaches have been taken, on one hand, extensive work on developing and studying effective field theories (EFT) \cite{Cepedello:2023yao,Criado:2021trs,Kundu:2021cmo}. While we recognize the power and usefulness of EFTs, they rely on the assumption that new physics are out of the production range of current experimental facilities, therefore they are not able to describe new physics effects that can not be integrated out. On the other hand, 
work has been done for comparing sets of models \cite{Belyaev:2021ngh,Belyaev:2022qnf}, our proposal fits well as a complementary approach to the one proposed in these references. 
We propose a complementary approach to the usual procedure in BSM phenomenology, that we dubbed as Model Benchmarking. The idea behind of model benchmarking is to group models with similar features and find observables that allow to discriminate between family members under the assumption of a BSM measurement. In order to illustrate the methodology, the following letter shows with an example how to perform a model benchmark analysis, from the definition of the model family to the computation of predictions.

This paper is structured as follows: In Section \ref{sec:model} we present the model setup and three seminal cases that we consider for this work, in Section \ref{sec:prod} we discuss the main processes for dark fermion production at CLIC, and show the cross section for the relevant processes. Section \ref{sec:kin} describes and characterizes the kinematical distributions for the different final states, whie additional discrimination criteria are given in Section \ref{sec:additional}. Finally, our conclusions are given in Section \ref{sec:conclusion}

\section{Model Family definition}\label{sec:model}

The first step of a model benchmark analysis consists on defining a criterion for model grouping. For the sake of this work, we focused on the scotogenic hypothesis, which proposes a common solution to the apparently independent problems of dark matter and neutrino mass generation. The first attempt to tackle these problems at the same time was the scotogenic model \cite{Ma:2006km}. Across the years, there have been many variations of this idea \cite{masses_and_mixings,scotosinglet,Singh:2023eye,Leite:2023gzl,Ahriche:2022bpx} , transforming the scotogenic model into the scotogenic paradigm. The common feature of these models is the interaction between particles from the dark sector and SM leptons. In what follows, we have considered the subgroup of models where dark matter arises from a dark fermion $N$, interacting with standard model leptons via a trilinear term involving a charged mediator. The helicity of the new fermion and the spin of the mediator determines the form of the interaction. For instance, consider the first scotogenic model \cite{Ma:2006km}, this model fulfills our criteria with a scalar mediator, $\eta$ and a right-handed Majorana fermion, $N_R$. For this model, the trilinear term has the following form:
\begin{equation}
    -\mathcal{L}_{scalar}=\beta \bar{L}\eta N_R+\text{h.c.}.
\end{equation}
On the other hand, the Vector Scotogenic Model \cite{dong2021,masses_and_mixings} also falls on our definition with a vector mediator, $V_\mu$ and a left-handed Majorana fermion $N_L$:

\begin{equation}
    -\mathcal{L}_{vector}=\beta \bar{L}\gamma^\mu V_\mu N_L +\text{h.c.},
\end{equation}

The consideration for models with Dirac fermions is not trivial, due to the fact that SM leptons transform under different representations of $SU(2)_L$. In order to build a simplified model with Dirac singlet fermions, we need two mediators, a $SU(2)_L$ multiplet ($X_L$) and a  singlet ($X_R$), the interaction lagrangian would have the following form:
\begin{equation}
    -\mathcal{L}_{dirac}=\beta_R \bar{l}_LX_L^- N_R +\beta_L \bar{l}_RX_R^- N_L +\text{h.c.}.
\end{equation}
Where we included only the component of $X_L$ that couples to charged leptons.
Many theories with extended scalar sectors present a similar structure. In most of them, there is a mixing inside the scalar sector:
\begin{equation}
    \begin{pmatrix}
    X_{1}\\X_{2}
    \end{pmatrix}=  \begin{pmatrix}\cos\theta &\sin\theta\\
    -\sin\theta &\cos\theta
    \end{pmatrix}   \begin{pmatrix}
    X_{L}^-\\X_{R}^-
    \end{pmatrix},
\end{equation}
where $\theta$ depends on the underlying theory. The interaction term for one of the physical scalars can be written as:
\begin{equation}
    -\mathcal{L}_{X_1}=\bar{l}  X_1[\beta_R\cos\theta P_R+\beta_L \sin\theta P_L ] N +\text{h.c.}.
\end{equation}

This interaction term has the same form that the slepton-lepton-neutralino term in the MSSM \cite{Rosiek:1995kg}. The form of the interaction depends entirely on $\beta_L,\beta_R$ and $\theta$. Depending on the choice of these parameters, it is possible to recover the previous term of the original scotogenic model. However, for the special case of $\tan\theta=\frac{\beta_R}{\beta_L}$ the interaction becomes non-chiral, making a clear difference between the previous models and the Dirac scenario. For the sake of phenomenological completeness, we are going to consider this very special case.
It's worth mentioning the similarities between this last model and the SUSY-inspired model presented in Ref. \cite{Baum:2020gjj}, in a broad sense, our proposal is a special case of the setup of this reference. It's this similarity that motivates the usage of SUSY-inspired language for referring to the physical fields involved in this setup. However, it's important to recall that the assumptions made for building the non-chiral interaction can be made for a variety of models beyond super symmetric scenarios. For instance, the scalar potential of lef-right symmetric models \cite{lr1,lr2,lr3} allows to build this type of term.

In order to compare these three models, we defined a unified labeling scheme for the free parameters relevant to our work. The relevant parts of the models are composed of two particles: $N$ and $X$ with masses $M_N$ and $M_{med}$, respectively. $N$ is coupled to SM leptons with an interaction mediated by $X$ and we are assuming that $X$ can couple with the SM Higgs boson (due to the fact  that most of these models are motivated by neutrino mass generation mechanisms).
Under this generic setup, the relevant lagrangian has the following form:

\begin{equation}
    \mathcal{L}_{dark}=\text{kinetic terms}-\lambda_{dark} |X|^2|\Phi|^2-\beta\mathcal{O}_{lXN}+\text{h.c.},
\end{equation}
With
\begin{equation}
    \mathcal{O}_{lXN}=
\begin{cases}
  \bar{l}_L X N_R, & \text{for scalar mediator} \\
   \bar{l}_L\gamma^\mu X_\mu N_L, & \text{for vector mediator}\\
  \bar{l} X N, & \text{for SUSY-inspired mediator.} 
\end{cases}
\end{equation}
By doing this parameterization, we make explicit the fact that we are dealing with isomorphic models, in the sense that they provide topologically equivalent vertices, but with different Lorentz and gauge structure. Due to this topological equivalence, the parameters in each model have the same meaning, allowing to make a direct comparison. 
Additionally, more complicated models can be accommodated by fixing reference values for the additional degrees of freedom.
The key point of this parameterization and labeling is that models take a secondary role: the important objects from the phenomenological perspective are the masses and couplings, and each model acts as a connection that generates an image of the parameter space into what we can call the observable space.
It's worth to note that we are keeping only one type of interaction between the mediator and the Higgs boson. In principle, there are other type of terms for mediators arising from $SU(2)_L$ doublets, however we neglect these terms for two reasons: on one hand, these additional terms don't arise for other representations of $SU(2)_L$, making it hard to compare models in the same parameter setting. On the other hand, additional interactions with the Higgs boson generate a mass splitting between $X$ and the other components of the multiplet. This mass splitting opens a decay channel that makes it hard to compare between models. Under this setup, the only decay for the mediator is governed by $\mathcal{O}_{lXN}$, opening a window for model discrimination using the mediator decay width (this will become evident in Section \ref{sec:additional}).
With the definition of the model's framework, we can find ways of distinguishing between these scenarios. Due to the main features of the model family, the CLIC experiment is a promising facility for probing leptophilic dark fermions. Therefore, in what follows we study how these models can produce distinctive signatures at CLIC.

\section{Production at CLIC}\label{sec:prod}

\begin{figure}[!h]
    \centering
    \includegraphics[width=0.6\textwidth]{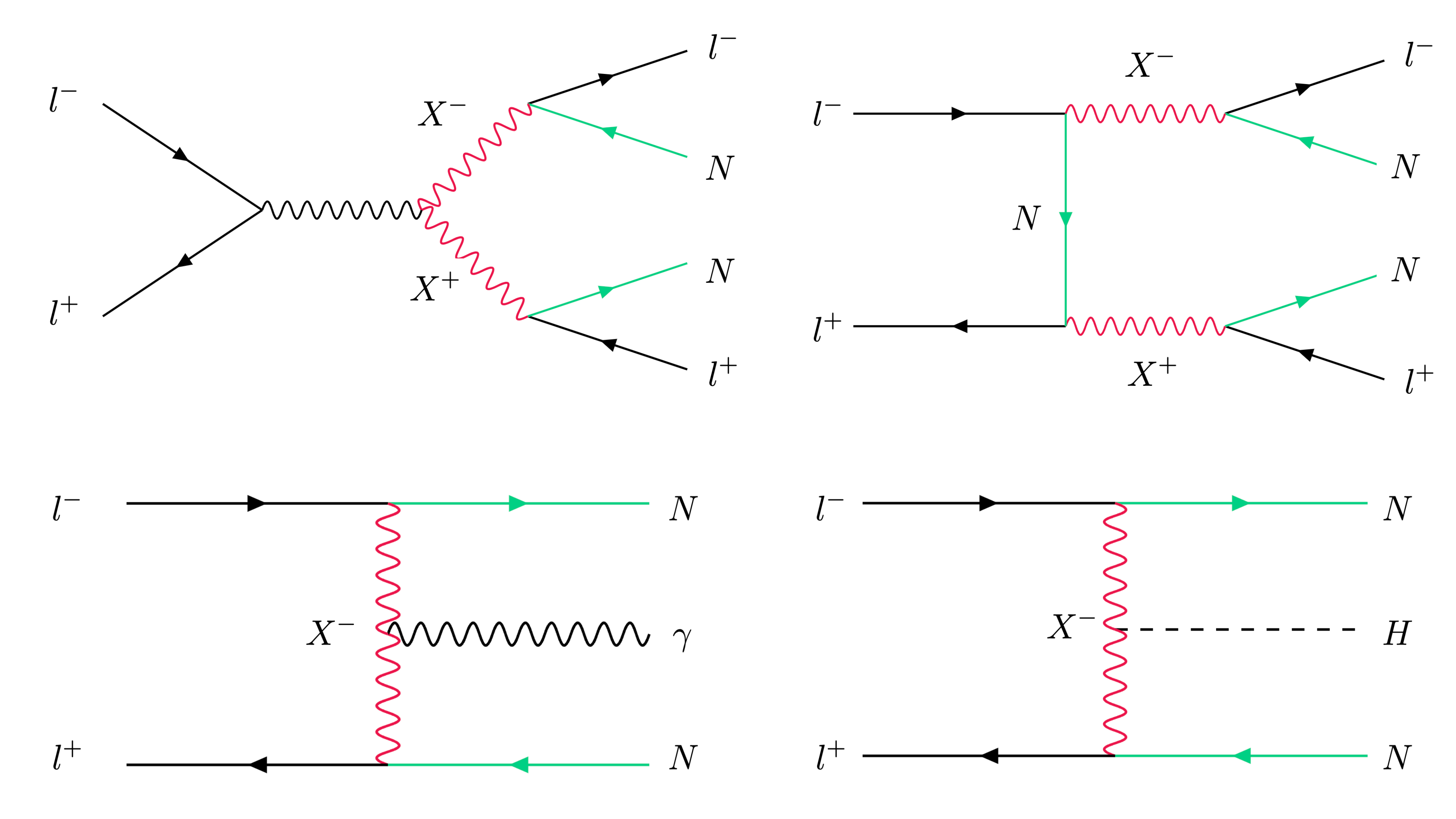}
    \caption{Feynman diagram for the production of dark fermions at lepton colliders}
    \label{diagrams}
\end{figure}

Typical searches for leptophilic dark matter models are focused on the $e^+ e^-  \slashed{E}_T$ final state \cite{Baumholzer:2019twf,scalar_scoto}. However, the structure of the model setup allows promising signals in the mono-photon and mono-Higgs final states. The relevant diagrams for all of these processes can be seen in Figure \ref{diagrams}. We implemented each model using Feynrules \cite{fr1,fr2} used Madgraph5\_aMC@NLO version 3.5.0  \cite{mg5} for the simulation of these processes at $\sqrt{s}=3$[TeV] considering unpolarized beams. The parameter values used for the simulation can be seen in Table \ref{params}. It's worth mentioning that the range for the mediator mass $M_{med}$ is related with unitarity constraints arising from the vector model \cite{vector_dm}.

\begin{table}[!h]
    \centering
    \begin{tabular}{|c|c|}
    \hline
    Parameter & Value \\
    \hline
    $\beta_e$     & $0.5$  \\
    $\beta_\mu$     & $0.0$  \\
    $\beta_\tau$     & $2.5$  \\
    $\lambda_{dark}$     & $5$  \\
    $\kappa$     & $-1$ \\
        $M_N$ & $50$[GeV]\\
    $M_{med}$ & $200-800$[GeV]\\
    \hline
    \end{tabular}
    \caption{Parameter setting for the simulations, note that $\kappa$ is a non-minimal gauge interaction that is present only in the vector mode, see Ref. \cite{vector_dm} for details.}
    \label{params}

\end{table}

\begin{figure}[!h]
    \centering
    \includegraphics[width=\textwidth]{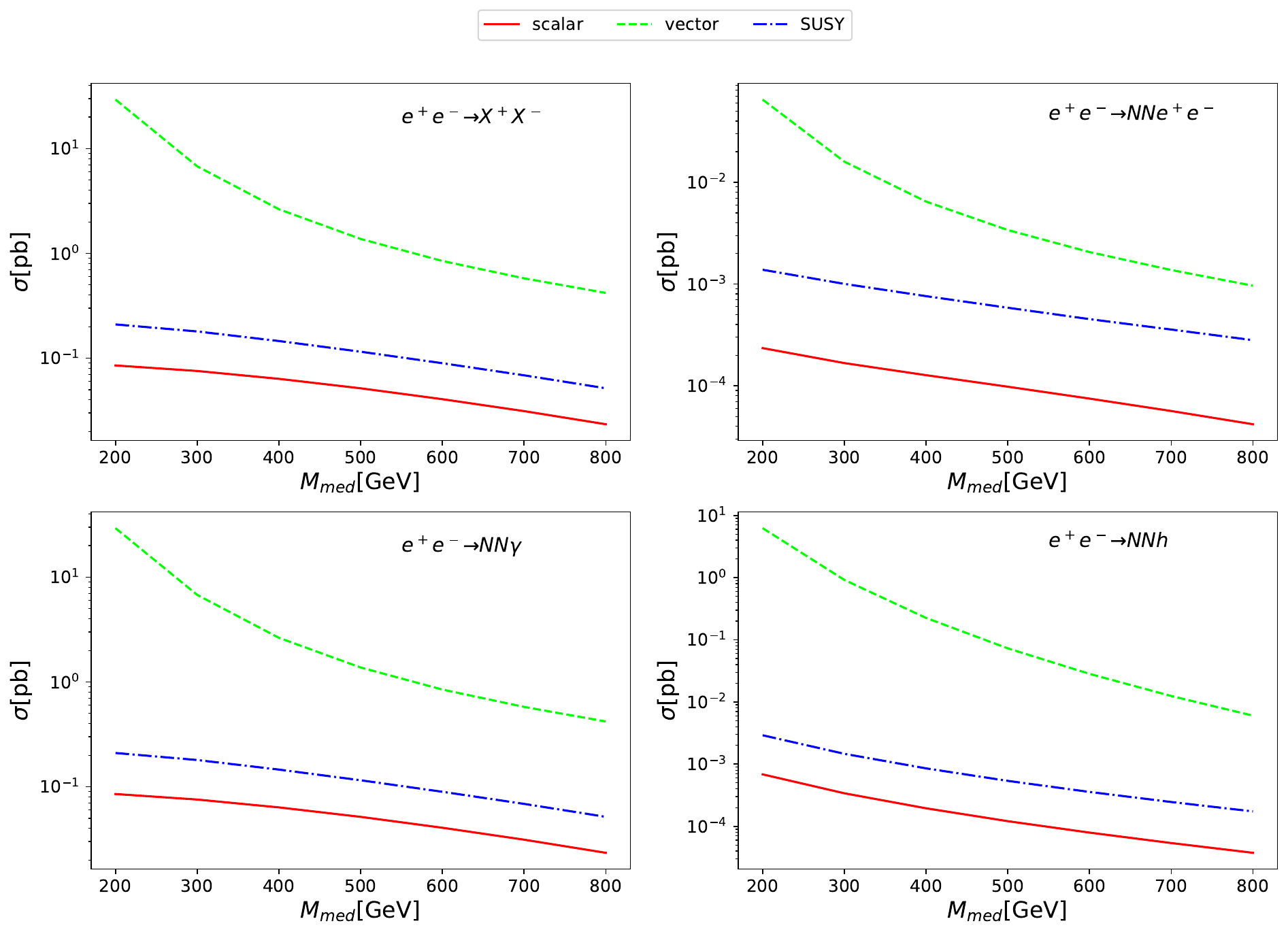}
    \caption{Parton level cross sections for the relevant processes regarding this simplified setup.}
    \label{crosses}
\end{figure}

The parton-level cross sections for the processes of interest can be seen in Figure \ref{crosses}. The vector model presents higher cross sections due to the contribution of polarization sum in the computation of the scattering amplitudes.

While lepton and photon reconstruction and tracking efficiencies are close to 1 \cite{CLICdp:2018vnx}, Higgs reconstruction is more complicated because it is an unstable particle. The reconstruction of a Higgs boson relies on the correct identification of the daughter particles. The most common decay corresponds to a $ b\bar{b}$ pair. Therefore, the effective cross section for mono-Higgs detection has the following form:
\begin{equation}
\sigma_{eff}=\epsilon \sigma_{parton} \text{BR}(H\to b\bar{b}),
\end{equation}
where $\epsilon $ stands for the reconstruction efficiency and $\text{BR}(H\to b\bar{b})$ is the branching fraction of the Higgs decaying to a $b\bar{b}$ pair. According to Ref. \cite{Abramowicz:2016zbo}, $\epsilon =0.35$ and $\text{BR}(H\to b\bar{b})=0.561$
the combined effect of these parameters shows that $\sigma_{eff}\approx 0.2 \sigma_{parton} $.
Cross sections by them selves are not enough for discriminating between models, therefore we studied the kinematics for each final state.
\section{kinematical analysis}\label{sec:kin}
As we stated in the previous section,
cross sections by themselves don't give a valid criterion for discriminating between models. Therefore, we studied the kinematics of these processes to define relevant criteria for signal classification.
\subsection{Di-lepton}
\begin{figure}[!h]
\centering
    \begin{subfigure}{0.45\textwidth}
    \includegraphics[width=\textwidth]{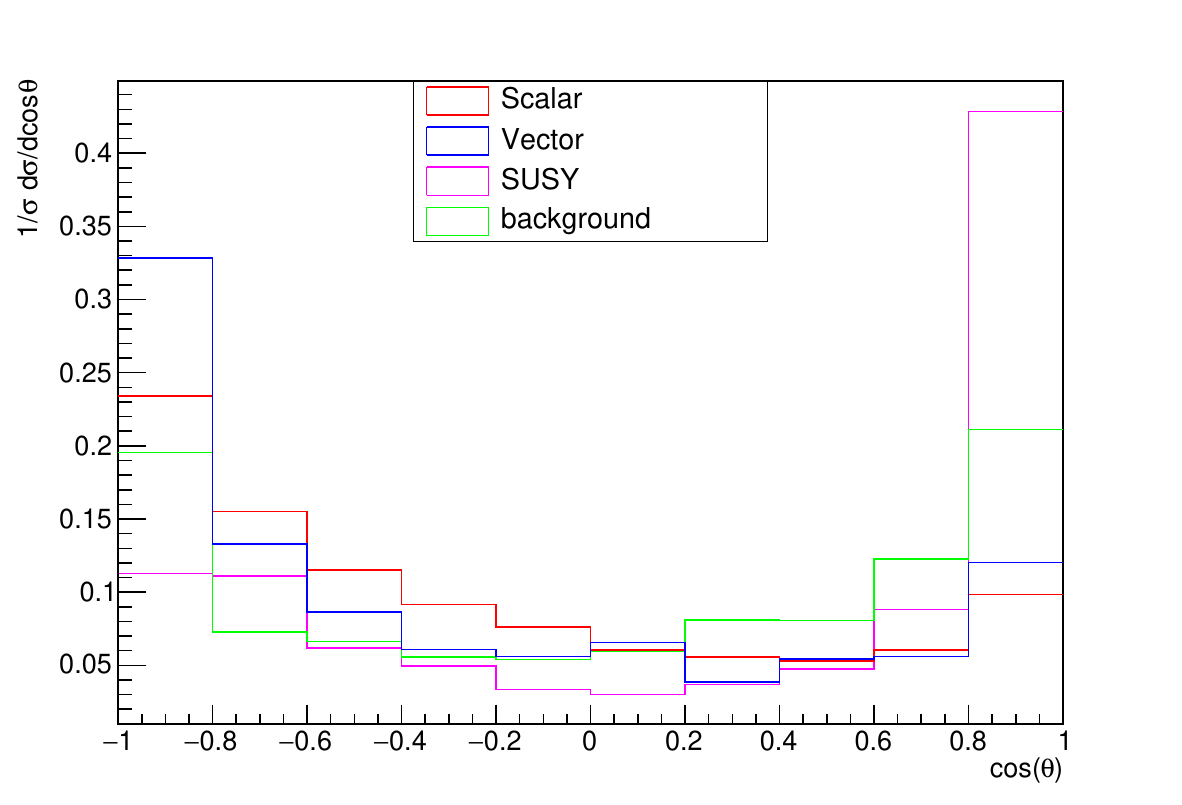}
    \caption{Dilepton angular distribution}
\end{subfigure}
\hspace{0.6cm}
    \begin{subfigure}{0.45\textwidth}
    \includegraphics[width=\textwidth]{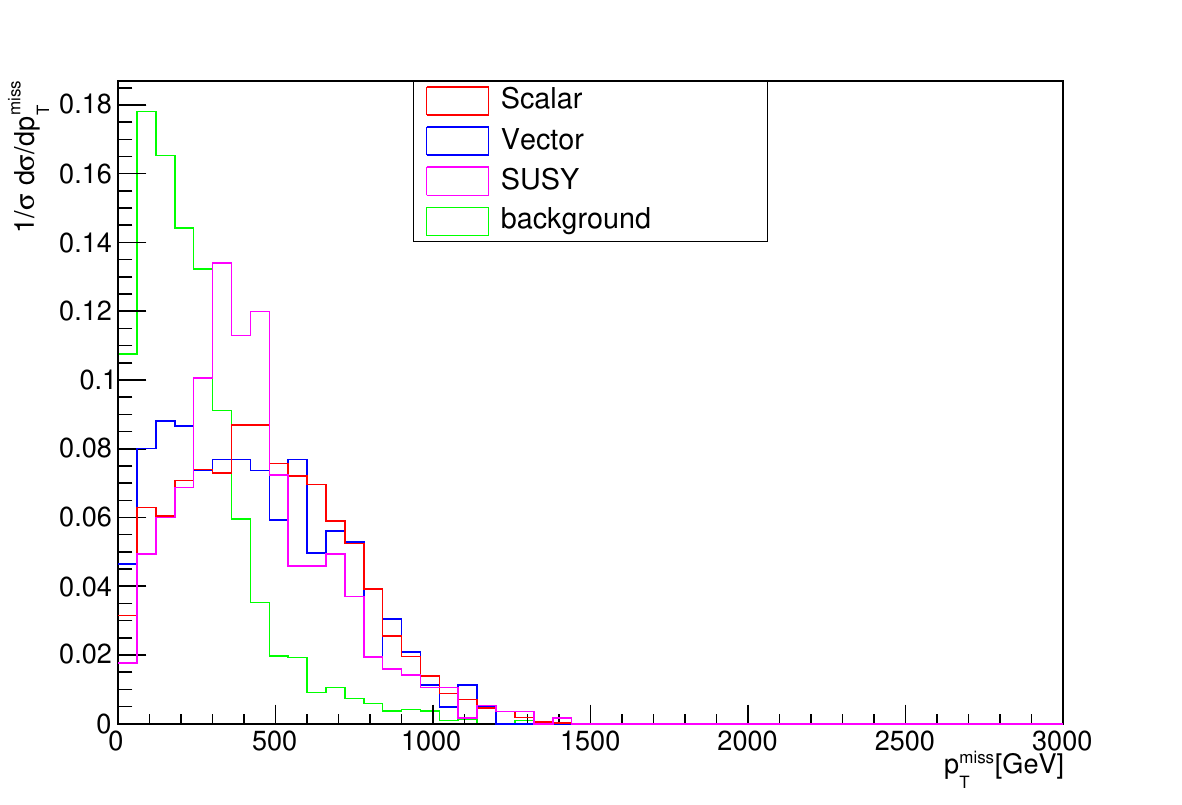}
    \caption{Missing transverse momentum}
\end{subfigure}

\caption{Kinematical distributions for the dilepton channel, for $M_{med}=800$[GeV]}
\label{kin_dilep}
\end{figure}

\begin{figure}[!h]
    \centering
    \includegraphics[width=0.45\textwidth]{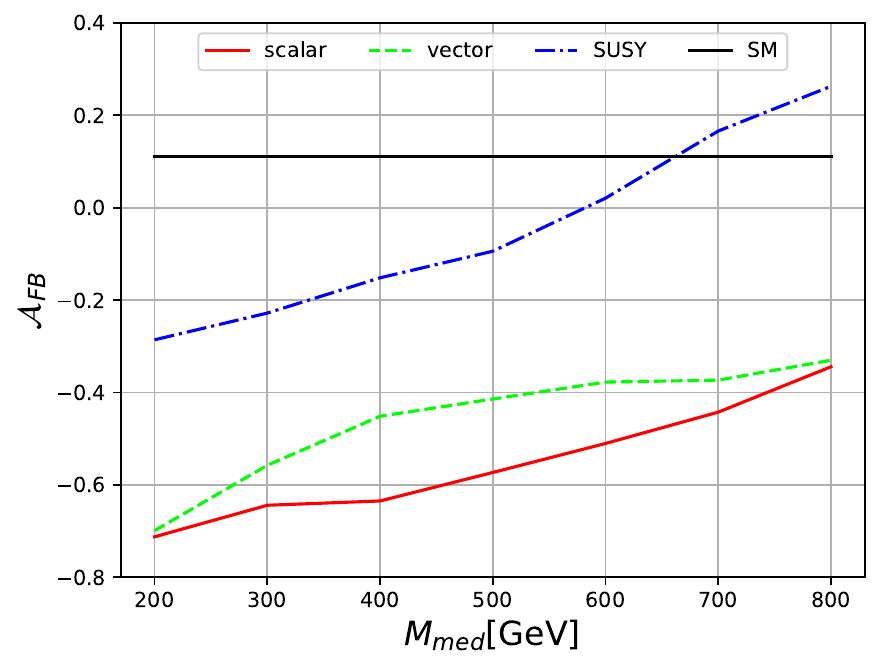}
    \caption{Forward-backward asymmetry for all the models}
    \label{afb}
\end{figure}

The dilepton channel  presents relatively small cross sections, however it shows a strong difference with the SM background. The relevant kinematical distributions for this channel are depicted in Figure \ref{kin_dilep}. In particular, the angular separation of the lepton pair is distinctive for every type of model. Due to this fact, we computed the forward-backward asymmetry, $\mathcal{A}_{FB}$ defined as:
\begin{equation}
\mathcal{A}_{FB}=\frac{\int_0^1\frac{\mathrm{d}\sigma}{\mathrm{d}\cos\theta}\mathrm{d}\cos\theta-\int_{-1}^0\frac{\mathrm{d}\sigma}{\mathrm{d}\cos\theta}\mathrm{d}\cos\theta}{\int_{-1}^1\frac{\mathrm{d}\sigma}{\mathrm{d}\cos\theta}\mathrm{d}\cos\theta}.
\end{equation}
We computed this quantity for each simulated sample. The result can be seen in Figure \ref{afb}. This observable is quite similar for the two chiral models, however, the non chiral interaction shows a clear difference. Moreover, there is a change of sign for a mediator mass of $M_{med}\sim 550$[GeV], showing a promising scenario for discriminating between models in this final state. 

\subsection{Mono-Photon}
\begin{figure}[!h]
\centering
    \begin{subfigure}{0.45\textwidth}
    \includegraphics[width=\textwidth]{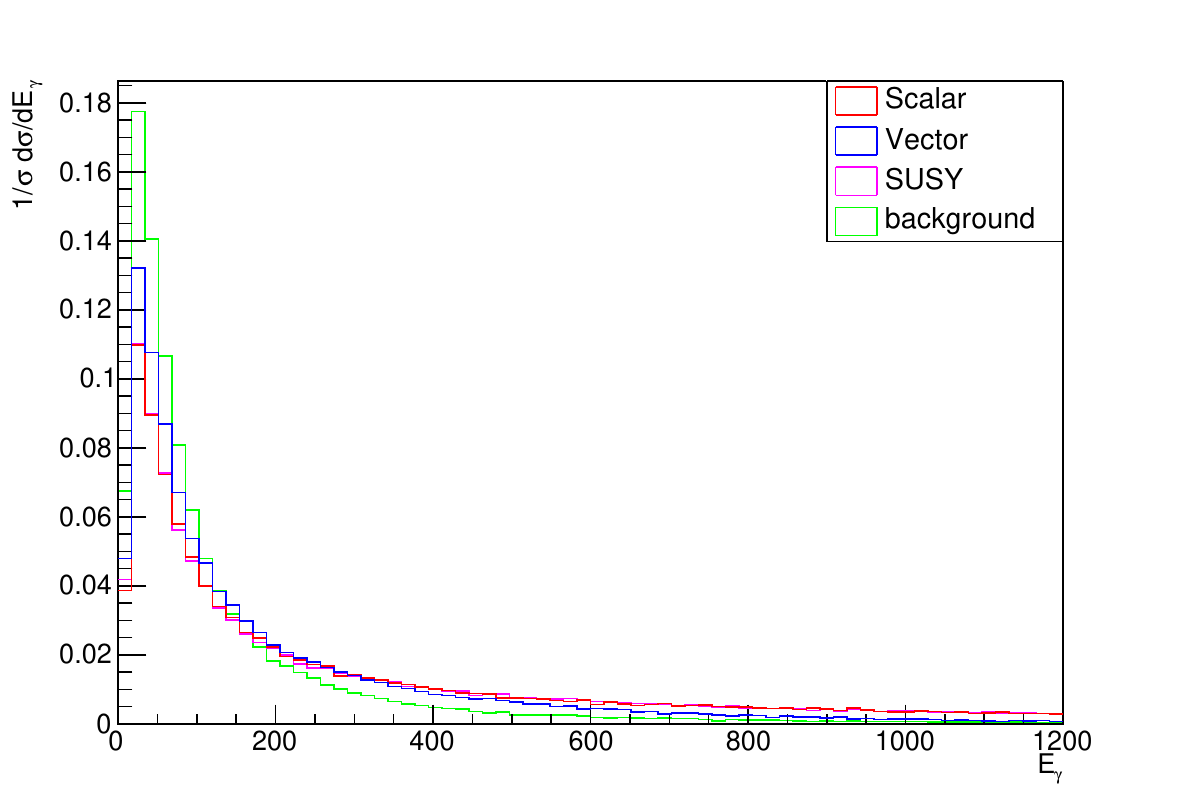}
    \caption{Photon energy}
\end{subfigure}
\hspace{0.6cm}
\begin{subfigure}{0.45\textwidth}
    \centering
    \includegraphics[width=\textwidth]{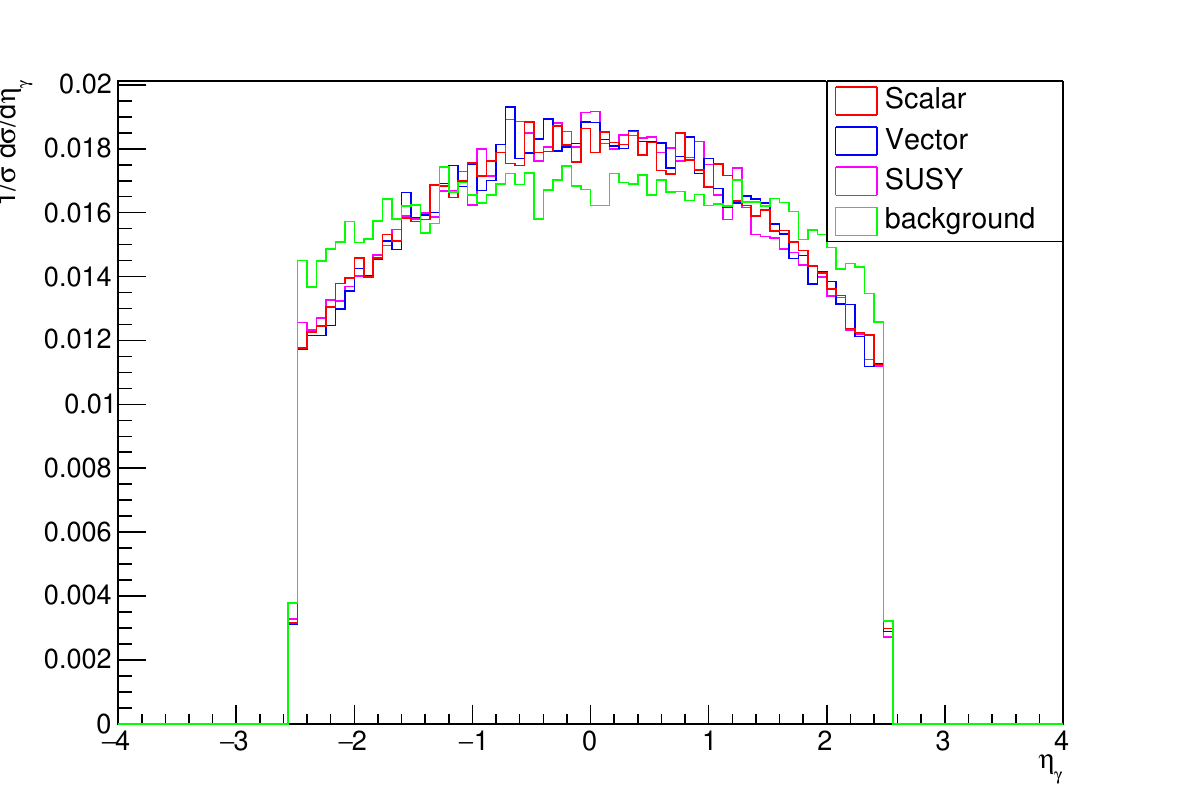}
    \caption{Photon pseudorapidity}
\end{subfigure}

\caption{Kinematical distributions for the mono-photon channel, for $M_{med}=800$[GeV]}
\label{kin_mono_a}
\end{figure}

This channel presents the stronger cross sections, however the kinematics of the process doesn't allow to make a clear discrimination between models,  as can be seen in Figure \ref{kin_mono_a}, the distributions are quite similar. The only noticeable difference is that the tail of energy distribution for the vector model is shortes than the other models, however, a realistic simulation considering detector effects will probably wash out this difference.

\subsection{Mono-Higgs}
\begin{figure}[!h]
\centering
    \begin{subfigure}{0.45\textwidth}
    \includegraphics[width=\textwidth]{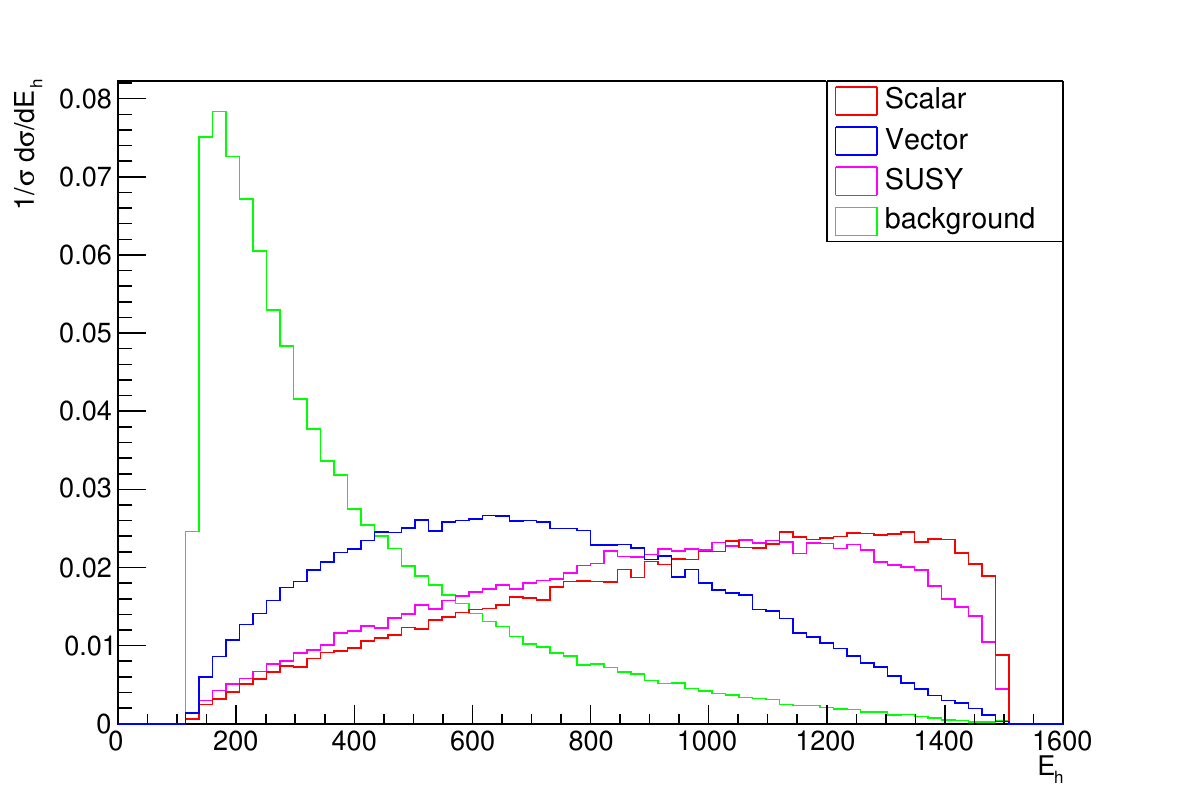}
    \caption{Higgs energy}
\end{subfigure}
\hspace{0.6cm}
\begin{subfigure}{0.45\textwidth}
    \centering
    \includegraphics[width=\textwidth]{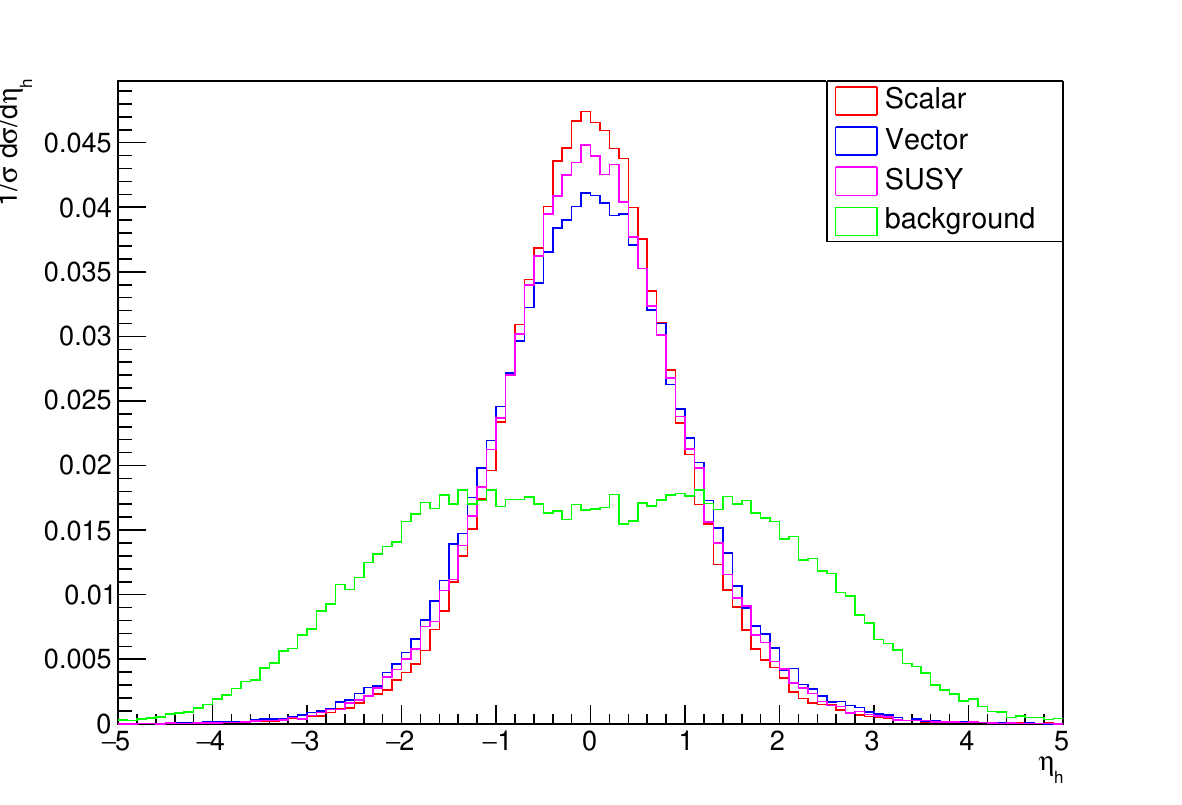}
    \caption{Higgs pseudorapidity}
\end{subfigure}

\caption{Kinematical distributions for the mono-photon channel, for $M_{med}=800$[GeV]}
\label{kin_mono_h}
\end{figure}

In contrast to the previous final state, the mono-Higgs signal presents notorious difference between the signals and the background (see Figure \ref{kin_mono_h}). On one hand, Higgs bosons produced in our models are notoriously more boosted in the transverse direction compared to the SM. On the other hand, the energy distribution of the Higgs boson is different in the vector model compared to the other two, opening a window for model discrimination.

\section{Additional discrimination criteria}\label{sec:additional}
\begin{figure}[!h]
    \centering
    \includegraphics[width=0.5\textwidth]{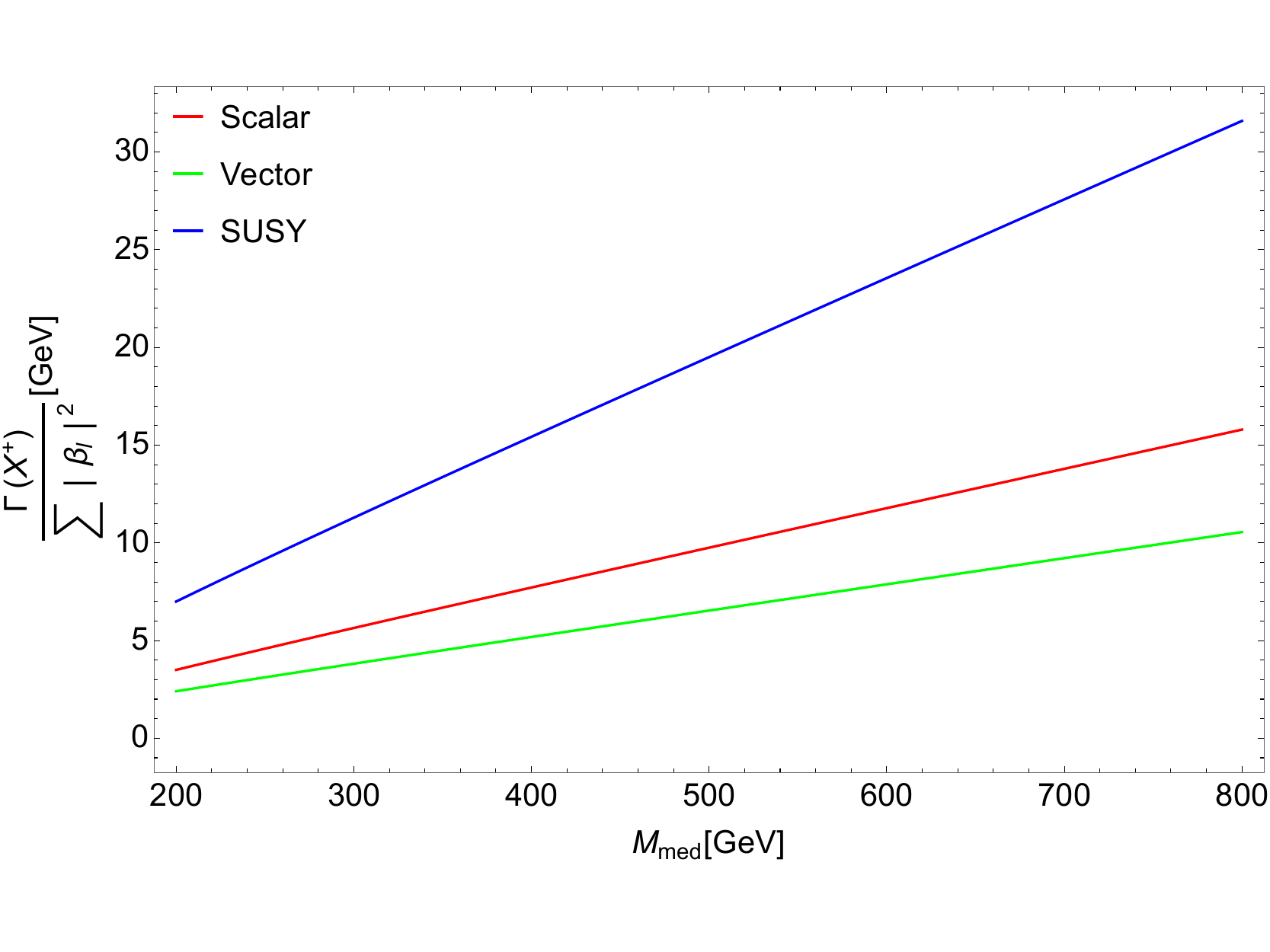}
    \caption{Mediator decay width for each model. It's worth to note that we factored out the coupling constants. In general, the order of magnitude of the decay widths can change, but the difference between them will be respected.}
    \label{widths}
\end{figure}

The kinematical analysis can be supplemented by considering different observables. One of them is the decay width of the mediator. Since the computation of the decay width depends on the integration phase space, it is sensitive to the Lorentz structure of the trilinear coupling. We computed the decay widths for each model (See Figure \ref{widths}). The SUSY-inspired model presents the largest decay width while the vector model presents the lowest. Precision tracking in the inner detectors can be used to measure the decay length of these charged particles. With this information, it can be easier to discriminate between models.

\begin{figure}[!h]
\centering
    \begin{subfigure}{0.45\textwidth}
    \includegraphics[width=\textwidth]{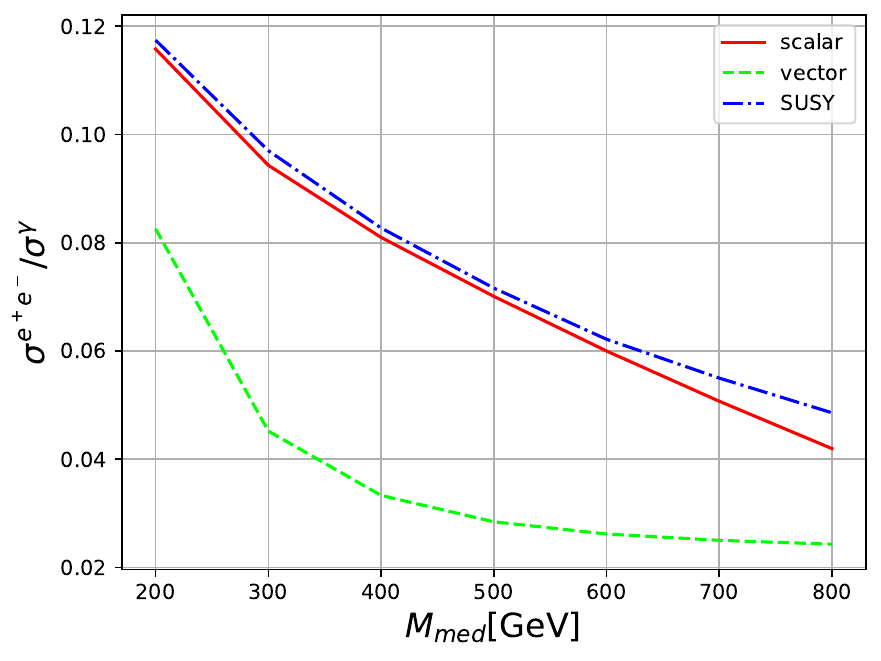}
    \caption{Mono-photon to dilepton ratio}
\end{subfigure}
\hspace{0.6cm}
\begin{subfigure}{0.45\textwidth}
    \centering
    \includegraphics[width=\textwidth]{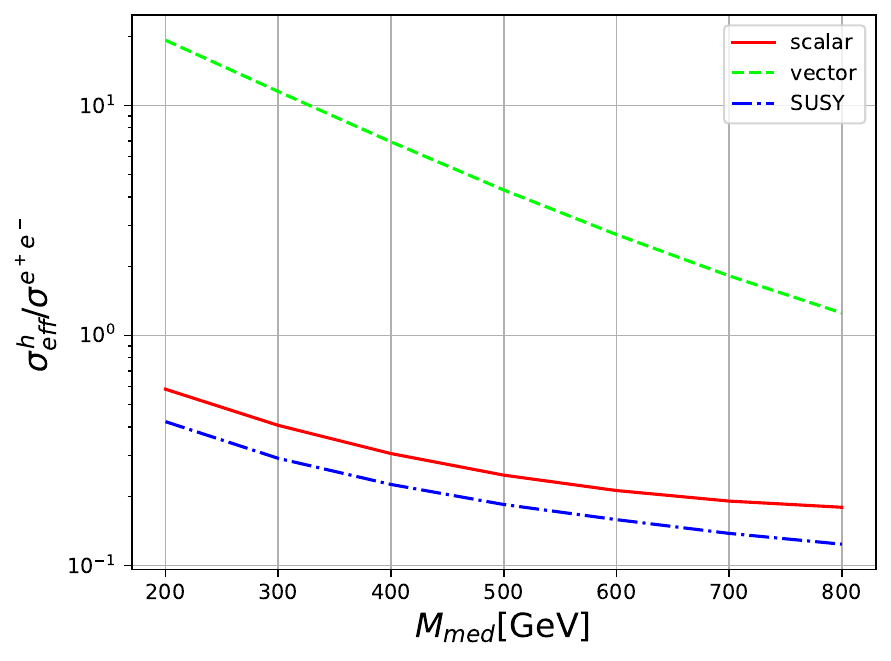}
    \caption{Mono-Higgs to dilepton ratio}
\end{subfigure}

    \begin{subfigure}{0.45\textwidth}
    \includegraphics[width=\textwidth]{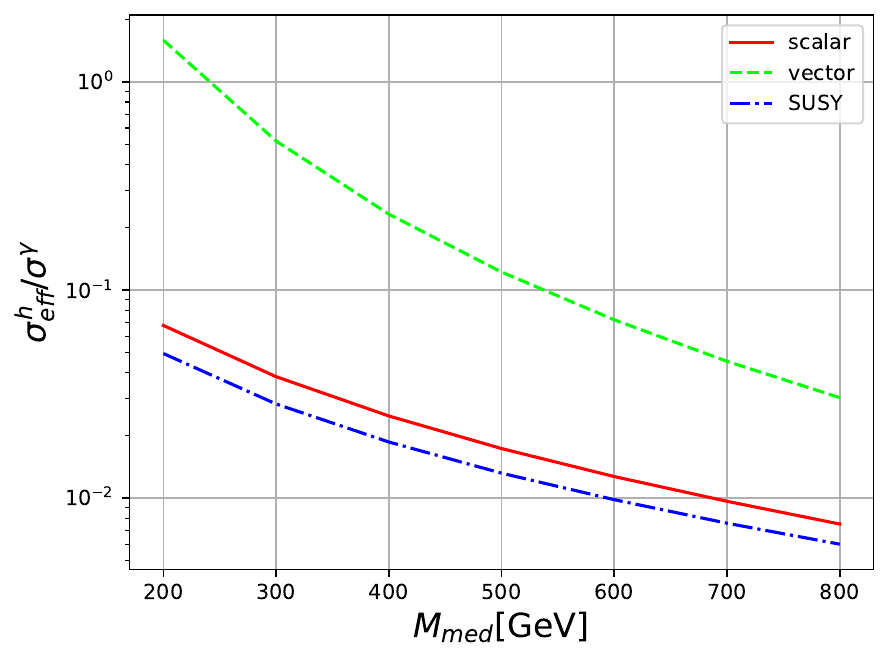}
    \caption{Mono-Higgs to Mono-photon ratio}
\end{subfigure}

\caption{Cross sections ratios for the models. The ratios were obtained considering the effective mono-Higgs cross section}
\label{cross_ratios}
\end{figure}

Another interesting observable is the ratio between cross sections for the given processes. As can be seen in Figure \ref{cross_ratios}, these ratios are rather unique for each model. These relationships between cross sections are translated into correlations in the number of events. This result shows that the discovery of any of these models requires to find excesses in all the three main final states. In order to show how these relationships could be a probe of these models, we consider predictions for an integrated luminosity of $\int\mathcal{L}\mathrm{d}t=2000[\text{fb}^{-1}]$ and $M_{med}=800$[GeV]. The expected significance for each model and final state can be seen in Table \ref{sig_bench}. It's important to recall that these values were obtained for a fixed parameter values, and this predictions should change. However, the relationship between the significance of different channels for the same model such prevail.

\begin{table}[!h]
    \centering
    \begin{tabular}{|c|c|c|c|}
    \hline
         & mono-photon &mono-Higgs &dilepton \\
         \hline
       Scalar  &   $0.83$    &      $0.034$     &    $0.15$     \\
       Vector  &    $32.62$   &      $5.52$     &     $3.38$    \\
       Susy  &     $4.78$  &       $0.16$    &      $0.99$   \\
       \hline
    \end{tabular}
    \caption{Expected significance for $\int\mathcal{L}\mathrm{d}t=2000[\text{fb}^{-1}]$ and $M_{med}=800$[GeV]. This table considers the corrected value for the mono-Higgs cross section.}
    \label{sig_bench}
\end{table}

\section{Conclusion}\label{sec:conclusion}
In this work, we introduced  model benchmark analysis for the study of isomorphic models, using as a proof of concept the family standard model extensions where a fermionic dark matter candidate couples to the lepton sector via an additional mediator. Under our framework, moodels can be classified by the nature of the dark fermions (Dirac or Majorana) and the spin of the mediator. We considered the double production of dark matter at CLIC on three different final states: the dilepton channel, the mono-Higgs and mono-photon. We studied the kinematical distributions for each channel for the search for distinctive patterns that can help to distinguish between models. Our results show that, while the mono-photon channel presents the higher cross section, there is a strong overlapping between the SM background and the different selected models. On the other hand, the mono-Higgs channel allows a clear differentiation between the signals and the background, and the signal is sensitive to the spin of the mediator. Finally, the dilepton channel presents the lower cross sections, but this channel allows to discriminate between the background and the three selected models, using the forward-backward asymmetry.

In addition to the kinematical analysis, we showed how these models can be separated by the measurement of the mediator lifetime. This observable plays a crucial role for the search of long living particles and can be used to distinguish between models.
Another key observable that can play a fundamental role for the discovery of new physics, is the correlation of events in the different final states, the number of events expected for each channel and each model is unique, and can give the stronger evidence for the discovery of new physics and the correct interpretation of the underlying theory.
We hope that the methodology proposed here can be useful for the analysis of different families of models and different experimental facilities, noting the importance of dedicating efforts to model benchmarking for future developments in particle physics phenomenology.
\section*{Aknowledgements}
This work was funded by ANID - Millennium Program - ICN2019\_044. AZ was partially supported
by Proyecto ANID PIA/APOYO AFB220004 (Chile) and Fondecyt 1230110.

\bibliographystyle{utphys}
\bibliography{References}
\addcontentsline{toc}{chapter}{\bibname}

\end{document}